\documentclass[two column,pra,showpacs, showkeys]{revtex4}

\usepackage{graphicx}
\usepackage{dcolumn}
\usepackage{bm}
\usepackage[latin1]{inputenc}
\usepackage{hyperref}
\usepackage{url}
\usepackage{amsmath}
\usepackage{epsfig}
\newcommand{\unit}[1]{\ensuremath{\, \mathrm{#1}}}

\begin{document}

\preprint{}

\title{Transition to quantum turbulence in a finite size superfluid}

\author{R.~F.~Shiozaki$^1$}\email{rfshiozaki@gmail.com}
\author{G.~D.~Telles$^1$}
\author{V.~I.~Yukalov$^2$}
\author{V.~S.~Bagnato$^1$}
\affiliation{$^1$Instituto de F\'{\i}sica de S\~{a}o Carlos, Universidade de S\~{a}o Paulo, C.P. 369, 13560-970 S\~{a}o Carlos, SP, Brazil \\ $^2$ Bogolubov Laboratory of Theoretical Physics, Joint Institute for Nuclear Research, Dubna, 141980, Russia}

\begin{abstract}

	A novel concept of quantum turbulence in finite size superfluids, such as trapped
bosonic atoms, is discussed. We have used an atomic $^{87}\mathrm{Rb}$
BEC to study the emergence of this phenomenon. In our experiment, the
transition to the quantum turbulent regime is characterized by a tangled vortex
lines formation, controlled by the amplitude and time duration of the excitation
produced by an external oscillating field. A simple model is suggested to account for
the experimental observations. The transition from the non-turbulent
to the turbulent regime is a rather gradual crossover. But it takes place in a
sharp enough way, allowing for the definition of an effective critical
line separating the regimes. Quantum turbulence emerging in a finite-size
superfluid may be a new idea helpful for revealing important features
associated to turbulence, a more general and broad phenomenon.

\end{abstract}



\maketitle

\section{Introduction}

Quantum Turbulence (QT) is a phenomenon related to the vortex dynamics in
superfluids and it can be realized in many different ways. In the case of
liquid helium, below the $\lambda$-point, moving grids and vibrating objects
can generate a fully tangled configuration of quantized vortices which
characterizes QT \cite{feyman,hall}. Within the context of low temperature
physics, QT has been studied since its discovery over fifty years ago
\cite{vinen}. One of the main motivations of studying QT is to establish
its relation with turbulence in classical fluids, where there is no
requirement for the vortex angular momentum quantization. For many years,
QT could only be studied in liquid helium ($^{4}$He and $^{3}$He)
\cite{donelly}. However, recently \cite{prl} Bose-Einstein Condensate
(BEC) of trapped gases has provided the main ingredients to study QT in
a more simple superfluids. The vortex nucleation in a BEC can be produced
by introducing a rotation in the trapped cloud \cite{rot1,rot2,rot3}. When
quantized vortices are generated, they arrange to form the Abrikosov lattices
\cite{shaeer}. These crystalline structures of quantized vortices result
from mutual interactions of the rotating field producing a repulsion between
vortex pairs which is balanced by the trapping potential. In such structures,
with a collection of vortices having the same direction and circulation
sign, QT cannot spontaneously occur. Since the main characteristic of QT
is a spatial tangled distribution of quantized vortices, one must have
vortex lines distributed in many spatial directions to reach such a
configuration. One method to achieve this was proposed by Kobayashi
and Tsubota \cite{michi}. In their proposal, combined rotations around
two orthogonal axes induce the nucleation of quantized vortex lines in
orthogonal directions with a clear evolution to QT. In a recent work
\cite{pra} a variant of the procedure suggested in Ref. \cite{michi} was
implemented. Henn \textit{et al.} \cite{pra} demonstrated that a special
type of oscillatory excitation imposed on a BEC generates vortices. When
vortices and anti-vortices are formed and proliferate through the sample,
the emergence of turbulence is observed \cite{prl} as a configuration of
tangled vortices. In fact, such oscillatory excitations generate
vortex-antivortex pairs \cite{threevortex}. The method of creating
vortices by means of oscillating external fields is a particular case
of the general method of creating coherent topological modes by such
oscillating fields \cite{YYB97,YYB02,YB09}.

Quantized vortices inside a condensed atomic cloud are the necessary
ingredients to produce quantum turbulence. Once the cloud is filled with
vortices and anti-vortices distributed in various directions, but not
arranged in a lattice, turbulence should naturally be established. The
production of vortices in a BEC is therefore the first step. The majority
of experimental groups have produced separate vortices and vortex arrays
by introducing a single-axis rotation in a BEC. The stirring technique
has been used with great success. In these experiments a similarity with
experiments performed with $^{4}$He \cite{turbulence} could be directly
observed. For large size BECs produced by the MIT group \cite{shaeer} a
large number of vortices with the same circulation could be produced and
configurations like Abrikosov lattices (originally observed for the
type-II superconductors) were clearly seen. As much as $130$ vortices
could be accommodated in the large MIT BEC. In such experiments, however,
no tangled configurations could be observed, but only crystalline
structures were produced.

In this communication, we analyze the process of generating QT in BEC,
focusing on the consequences of having a finite-size sample which
constitutes an important property of superfluids originating from cold
trapped atoms. In this case, the formed vortices spread inside the trap
until turbulence is developed. We analyze this effect in terms of the
amplitude and time of excitation by the oscillatory perturbation. We
suggest a simple model demonstrating the existence of an effective
transition line between non-turbulent and turbulent regimes. A comparison
with experimental data is presented, showing a good qualitative agreement.
We start with a brief description of the emergence of QT in a BEC,
followed by a general analysis and a comparison with experimental
observations. Finally, arguments are presented concerning the importance
of the phenomenon of quantum turbulence in finite superfluids.

\section{Experimental observation}

The vortex nucleation can be done by imposing an external oscillating field
on the condensed trapped atomic cloud, producing an effect equivalent to rotation.
Indeed, those oscillations generate vortex-antivortex pairs \cite{YYB97,YYB02,YB09}.
Vortex nucleation is expected to take place when the perturbation provides
enough energy.

Our system is composed of $^{87}$Rb atoms forming a BEC described earlier
\cite{prl,pra}. We used a combination of oscillations introduced by the
superposition of coils to the conventional trapping field. The anisotropic harmonic 
potential of a quadrupole-Ioffe-configuration (QUIC) trap is approximately given 
by $V=\frac{1}{2} m (\omega_{\rho}^2 \rho^2 + \omega_z^2 z^2)$.
In our case, $\omega_{\rho} / \omega_{z} = 9$, which corresponds to a cigar
shaped trap. Superimposing off-axis coils on this trap modifies the potential
so that the long axis is rotated while the minimum is displaced. This type
of excitation can produce different numbers of vortices, depending on the
oscillation amplitude and excitation time \cite{pra}. This type of
excitation should create both vortices and anti-vortices. Observations of
three vortices inside an atomic cloud provide good evidence for the existence
of vortex and anti-vortex  pairs, as discussed elsewhere \cite{threevortex}. Since 
the vortex and anti-vortex pairs are not exactly parallel to each other (oscillations
are not confined to a plane) the vortex and anti-vortex are expected to live longer 
than if they would be parallel. We have fixed the excitation frequency of $200\unit{Hz}$ and 
considered a range of amplitude and time intervals of the applied external field. 
The results are observed after a time-of-flight (TOF) of $20\unit{ms}$. The overall cloud characteristics are determined by the trapping harmonic potential, whose frequencies 
are $f_z=23\unit{Hz}$ and $f_{\rho}=207\unit{Hz}$. The cigar shaped BEC contains about 
$2\times10^5$ atoms with a typical condensate fraction ranging from $40$ to $70\%$.

\begin{figure}
\centering
 \includegraphics[scale=0.6]{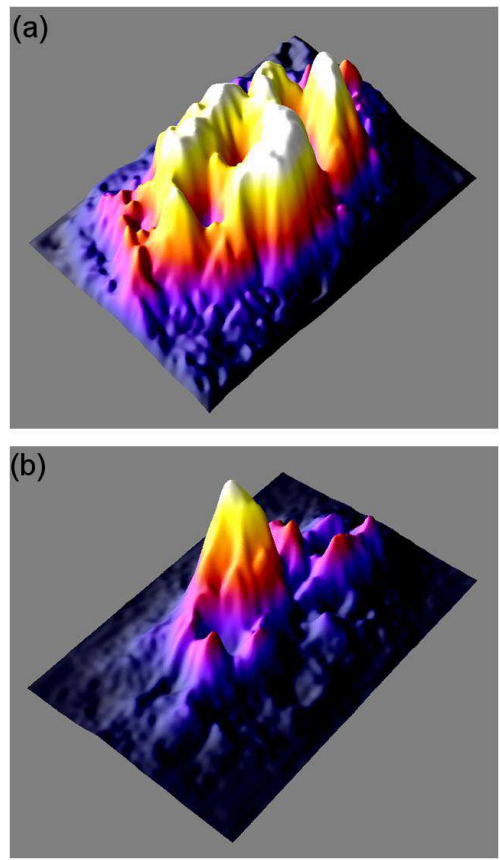}
 \caption{Atomic optical density images of: (a) non-turbulent cloud with
well-defined separate vortices and (b) turbulent cloud, where the partial
absorption changes along the image due to the existence of tangled vortices. The
images were taken after 15 ms of free expansion.}
\label{fig:clouds3d}
\end{figure}

As the oscillation amplitude/time increases, the vortices start to be
nucleated in agreement with  Ref. \cite{pra}. The number of vortices in the cloud is
counted as the clearly distinguishable dark regions in the absorption image
after TOF and it changes with the amplitude and time interval of the excitation
\cite{diagram}. We have varied the amplitude from $0$ to $200\unit{mG/cm}$ and the
excitation time up to $60\unit{ms}$. We are interested in studying the crossover region
from the observation of vortices to the turbulent cloud. The region in the diagram,
separating both regimes, as well as the typical images characterizing the
so-called non-turbulent and turbulent regimes are shown in
Fig.\ref{fig:clouds3d} and Fig.\ref{fig:diagram}.

In order to distinguish between the non-turbulent and turbulent regimes, we
have adopted the following criterion. In Fig.\ref{fig:clouds3d}(a), we
present a density profile, where one can clearly see the vortices as the
dark regions (valleys) spread inside the cloud. In this case, the typical
Thomas-Fermi aspect ratio inversion is observed during the TOF (free fall), as
expected in a \textit{non-turbulent} gas. On the other hand, the turbulent
regime is characterized by tangled vortex lines spread all over the cloud.
As a result, the dark region contrast essentially fades away, as is seen in 
the absorption images of Fig.\ref{fig:clouds3d}(b). Besides that, the behavior 
of a QT cloud during the TOF is different than that of a regular BEC. The axes 
aspect ratio is kept essentially unchanged \cite{prl}. The tangled vortices 
make the whole system isotropic \cite{Y10}. Combining these characteristics, 
we define a \textit{turbulent} cloud. This is the criterion for distinguishing 
between the turbulent and non-turbulent regime.

\section{Theoretical model}

\begin{figure}
\centering
 \includegraphics[scale=0.31]{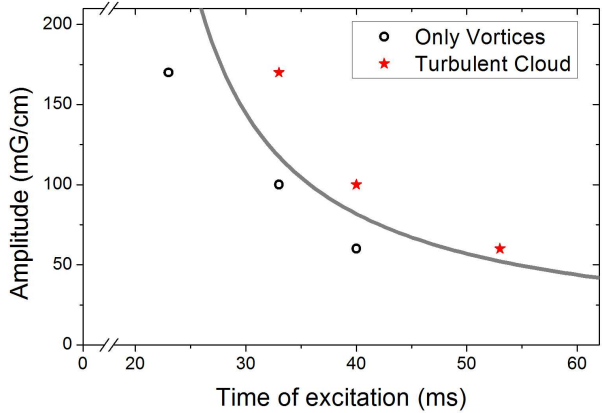}
 \caption{Diagram, on the excitation amplitude-time plane, demonstrating the crossover between the non-turbulent atomic cloud, with well-defined separate vortices, and the turbulent cloud, with tangled vortices. The plotted line is based on Eq. (\ref{eq:Acrit}) considering halfway points with the same amplitude as experimental critical points.}
\label{fig:diagram}
\end{figure}

According to Fig.\ref{fig:diagram}, there exists a well-defined parameter region,
where the non-turbulent regime evolves to the turbulent regime. To understand
the observed behavior, we propose a simple model based on energy-balance arguments,
which are quite general and do not depend on the actual mechanism type
needed for the vortex creation or its dynamics inside the cloud. The
nucleation of vortices is due to the instability of collective excitations
arising from the energy pumped by external oscillating perturbations \cite{Y10,orsay}.

We start by noting that there should exist a certain energy amount that is necessary 
to pump into the superfluid atomic cloud for the vortex formation. Following \cite{pethick}, 
we write down the energy needed for the vortex nucleation as
\begin{equation}\label{eq:Evort}
E_{vort}=\frac{\hbar^2}{m l_0^2} \ln{\frac{l_0}{\xi}} \; ,
\end{equation}
where, $\xi=(\sqrt{8 \pi n a_s})^{-1}$ is the healing length, $n$ is the BEC peak density, 
$a_s$ is the s-wave scattering length, and $l_0$ is the vortex line length. We assume that 
the latter is approximately equal to the effective cloud's harmonic oscillator length, 
\begin{equation}\label{eq:lzero}
l_0=a_{ho}= \sqrt{\frac{\hbar}{m (\omega_{\rho}^2 \omega_z)^{1/3}}} \; .
\end{equation}

Then, let $R_{pump}$ be the rate of the total energy pumped into the cloud by the 
external oscillating field, and $\eta$ be the energy fraction converted to rotation. 
Therefore the total energy needed for the vortex formation can be written as
\begin{equation}\label{eq:Ezero}
E_{pump}=\eta R_{pump} (t-t_0) \; ,
\end{equation}
where $t$ is the elapsed excitation time and $t_0$ is the minimal time period needed 
for the first vortex creation. The first vortex is created when $E_{pump} \approx E_{vort}$. 
If after the time $t$ of pumping, the number of vortices $N_{vort}$ is formed, then 
the energy balance implies that
\begin{equation}\label{eq:best}
\eta R_{pump} (t-t_0) = N_{vort} E_{vort} \; ,
\end{equation}
and the expected number of vortices is
\begin{equation}\label{eq:number}
N_{vort} = \frac{\eta R_{pump}}{E_{vort}} (t-t_0) \; .
\end{equation}

This gives us a good estimate for the number of vortices observed in the atomic 
cloud, as a function of the excitation elapsed time, which can be compared to the 
time dependence reported in \cite{diagram}. Here, we have not explicitly taken 
into account the vortex-antivortex annihilation, but this effect could be 
incorporated into the value of the coefficient $\eta$. This estimate depends 
neither on the cloud motion during the excitation, nor on the presence of 
collective modes that certainly arise \cite{icap}. Assuming that the turbulence 
onset takes place when the atomic cloud is heavily populated by vortices, 
and $N_{vort} \xi$ is of the order of the characteristic trap size, we conclude 
that turbulence should arise when the number of vortices is 
$$
N_{vort} \approx \frac{l_0}{\xi} \; .
$$
Near the frontier between the two regimes, the above equations result in the 
critical behavior of $R_{pump}$, generating turbulence:
\begin{equation}\label{eq:Rcrit}
R_{pump}= \frac{l_0E_{vort}}{\xi \eta (t-t_0)} \; .
\end{equation}

The energy pump rate, $R_{pump}$, is proportional to the ratio between the 
external field amplitude, $A$, and the oscillating (pump) frequency. Therefore, 
the critical pumping amplitude can be expressed as
\begin{equation}\label{eq:Acrit}
A_c(t) = \frac{C}{t-t_0} \; .
\end{equation}

The borderline separating the non-turbulent and the turbulent regimes is 
shown in Fig.\ref{fig:diagram}. For our experimental system, we have found 
$C \approx 1.6 \unit{ms(G/cm)}$,  $t_0 \approx 17\unit{ms}$, $\xi \approx 0.06 \unit{\mu m}$, 
$l_0 \approx 1.08 \unit{\mu m}$, and $E_{vort} \approx 20 \unit{nK\times k_B}$, as the 
characteristic values. With these quantities, we expect $N_{vort}\approx 20$ 
for the QT onset, which is in good agreement with the recent work \cite{diagram}, 
where the turbulence is observed when the number of vortices is close to the 
value $20$ found above.

\section{Role of the finite size of atomic systems}

The existence of the critical borderline, separating the non-turbulent and 
turbulent regions in the diagram of Fig.\ref{fig:diagram}, is closely related to 
the finite-size of the trapped superfluid system, whose effective size is given 
by the characteristic anisotropic harmonic trap length $l_0$. Quantum turbulence 
develops when the trapped atomic cloud becomes densely filled with vortices. Then 
the energy, pumped into the system, transforms not only into the newly formed 
vortices, but also to their rapid motion, with the formation of their tangled 
distribution, accompanied by the appearance of reconnections and the formation 
of Kelvin waves. At this point, the absorption images become hazy, which is a 
manifestation of the arising turbulence \cite{feshbach}.

The standard experiments with $^{4}$He and $^{3}$He do not fulfill the conditions, 
where finite-size effects would become crucially important. Hence, the superfluid 
clouds, formed by trapped atomic BECs, represent a whole new class of systems, 
whose properties essentially depend on finite-size effects, as well as on 
interactions.

Effects, related to finite temperature, should also be important, though, at this 
time, we do not have enough data for performing the corresponding analysis. 
The pumped energy is, certainly, partially transformed into the thermal cloud, which 
is necessary to take into account for a more detailed consideration \cite{Y09}. 
It can also be that the losses, caused by the thermal cloud, could be responsible 
for the value of the delay time $t_0$. The existence of the critical line in 
Fig.\ref{fig:diagram} requires that $t$ be larger than $t_0$. In order to generate 
vortices at $t \approx t_0$, one needs a very large amplitude. But a strong or long 
pumping should produce a large admixture of the thermal fraction \cite{Y10}.

\section{Conclusions}

Summarizing, we have presented experimental data and offered an explanation for 
the transition between a non-turbulent and turbulent regimes observed in a superfluid 
formed by a Bose-Einstein condensate of trapped atoms. The transition region occurs 
on the amplitude-time parameter of the related excitation. The character of the 
observed transition region is closely connected with the finiteness of the superfluid, 
since quantum turbulence arises when the sample becomes densely saturated with vortex 
lines. A simple model allows us to qualitatively understand the observed main features 
of the phenomenon.

Further studies, both experimental and theoretical, are necessary for the better 
understanding of this phenomenon of quantum turbulence in finite-size superfluids.
This concept of quantum turbulence in finite systems can also be applied to small 
$^{4}$He droplets, though their experimental realization can be much more complicated 
than the creation of atomic clouds in traps. The possible existence of finite-size 
effects in turbulent superfluids opens a novel direction in the investigation of 
finite systems, such as trapped atoms and liquid droplets.

At low temperatures, fermions, depending on the sign of their interactions, can form
either superfluid molecular BEC or paired superconductor-type fluid, both of which
can exhibit superfluid properties \cite{Ket08,Sal09}. It is, therefore, feasible to produce 
quantum turbulence in such systems and to observe finite-size effects in fermionic
turbulent fluids, similar to those observed in trapped bosonic superfluids.

\section{Acknowledgements}

We appreciate collaboration with E. Henn, J.A. Seman, G. Roati, K. Magalh\~aes,
F. Poveda-Cuevas, S. Muniz, M. Kobayashi, K. Kasamatsu, and M. Tsubota. This work 
was supported by FAPESP and CNPq. One of the authors (V.I.Y.) acknowledges financial 
support from the Russian Foundation for Basic Research.

\end{document}